\documentclass{aa}  

\usepackage{graphicx}
\usepackage[varg]{txfonts}
\usepackage[colorlinks=true, linkcolor=blue, citecolor=blue, filecolor=blue, urlcolor=blue]{hyperref}

\newcommand{\rsun}{$R_{\odot}$}
\newcommand{\msun}{$M_{\odot}$}

\begin{document}

   \title{\textit{Swift} X-Ray and UV Observations of six \textit{Gaia} Binaries supposedly containing a Neutron Star.}

    \titlerunning{\textit{Swift} observations of six Neutron Star binary candidates}
   \author{ B.~Sbarufatti\inst{1}
     \and F.~Coti Zelati\inst{2,3} 
     \and A.~Marino\inst{2,3}
     \and S.~Mereghetti\inst{4}
     \and N.~Rea\inst{2,3}
     \and A.~Treves\inst{5}
   }
   \institute{
     INAF–Osservatorio Astronomico di Brera, Via E. Bianchi 46, Merate (LC), I-23807, Italy
     \and
     Institute of Space Sciences (ICE, CSIC), Campus UAB, Carrer de Can Magrans s/n, E-08193 Barcelona, Spain
     \and
     Institut d'Estudis Espacials de Catalunya (IEEC), 08860 Castelldefels (Barcelona), Spain 
     \and
     INAF-Istituto di Astrofisica Spaziale e Fisica Cosmica di Milano, via A. Corti 12, I-20133 Milano, Italy
     \and
     Universit\`a dell'Insubria, Dipartimento di Scienza e Alta Tecnologia, Via Valleggio 11, I-22100 Como, Italy
             }

\abstract{ Recent observations have led to the discovery of numerous optically selected binaries containing a undetected component with mass consistent with a compact object (neutron star or white dwarf).  Using the the \textit{Neil Gehrels Swift Observatory} we have carried out X-ray and UV observations of a small sample of these binaries. Four systems are wide (with orbital period P$>300~d$), and they were chosen because of their small distance (d$<$250 pc) and the mass of the collapsed component favoring a neutron star. Two other are compact systems (P$<0.9~d$), with convincing evidence of containing a neutron star. The source 2MASS J15274848$+$3536572 was detected in the X-ray band, with a flux of $5\times10^{-13}$ erg~cm$^{-2}$~ s$^{-1}$ and a spectrum well fitted by a power law or a thermal plasma emission model. 
This source also showed an UV (2200 $\AA$) excess, which might indicate the presence of mass accretion.
For the other targets we derived  X-ray flux upper limits of the order of $10^{-13}$ erg~cm$^{-2}$~s$^{-1}$ . 
These results  are consistent with the hypothesis that the collapsed component in these six systems are neutron stars.}  

   \keywords{
   Stars: neutron,
   X-rays: binaries
               }
   \maketitle

\section{Introduction}

Since the 1960s, it was clear that neutron stars (NSs) and black holes (BHs) in  binary systems could be detected through the orbital motion induced on their companion stars (e.g., \citet{Guseinov1966}). However, these compact stars were first found in different ways: NSs were discovered as radio pulsars \citep{Hewish1968}, while stellar-mass BHs were identified through observations of accretion-powered X-ray binaries \citep{Bolton1972,Webster1972}.  
The typical steps that led to the identification of most known stellar-mass BHs were the discovery of luminous (transient) X-ray sources, the identification of their optical counterparts, and subsequent spectroscopic observations to determine the binary parameters, in particular to constrain the mass of the compact object.
Currently a few tens of stellar-mass BHs in X-ray binaries have been identified in this way \citep[e.g.][]{Corral2016}.
    
The appearance of large stellar surveys in recent years has enabled to search systematically for binary systems, with only one visible component, and a companion whose mass is compatible with that expected for NSs or BHs. Particularly relevant in this regard is the contribution of the \textit{Gaia} mission \citep{gaia}, which has covered $\sim~10^9$ Galactic stars for several years. Not only can \textit{Gaia} identify binary systems through repeated multiband photometry, but it also has superb astrometrical capabilities that provide precise distance information (at least for relatively bright sources). In the case of wide binaries, astrometry alone allows to detect and study the dynamics of the binaries. 
Although, the primary source of data for these analyses comes from optical observations, multi-wavelength follow-up can be useful to confirm the presence of compact objects in these binaries.

In this paper we present X-ray and UV observations obtained with the \textit{Neil Gehrels Swift Observatory} of a few  binaries selected from optical observations as strong candidates for containing a NS. Four targets were chosen because they are those with the smallest distance among the candidates found in a sample of astrometric binaries \citep{Andrews2022}. They have long orbital periods ($\gtrsim1$ yr), high eccentricities, and are closer than 250 pc. The other two targets are short period binaries (P$< 1~d$) for which the masses were determined from photometric and spectroscopic observations \citep{Yuan2023, Lin2023}. 
The non-collapsed components in all these binaries are main sequence stars.
The main properties of our targets are summarized in Table \ref{table:1}, where we also indicate the short source names adopted in this paper. 
   
    \begin{sidewaystable*}
        \caption{Source Properties}
        \label{table:1}
        \centering
        \begin{tabular}{lllllllll} 
        \hline\hline 
        Source Name & \textit{DR3 Gaia} ID & distance & period\tablefootmark{1} & eccentricity & $m_1$\tablefootmark{2} & Type\tablefootmark{3} & G& $m_2$\tablefootmark{4}  \\
      &   & \multicolumn{1}{l}{pc}  & \multicolumn{1}{l}{d}  & \multicolumn{1}{l}{}  &\multicolumn{1}{l}{\msun} & \multicolumn{1}{l}{} & \multicolumn{1}{l}{mag}& \multicolumn{1}{l}{\msun} \\
        \hline
    0616+2319     &  3425175331243738240 & 1111$\pm$30    & 0.866605$\pm 1.0\cdot10^{-6}$     & 0.0007$^{+0.007}_{-0.0005}$& 1.7$\pm$0.1    & G7 & 13.21& 1.1-1.3    \\
    1220+5841     &  1581117310088807552 & 219$\pm$3    & 927$\pm11$        & 0.52$\pm$0.01 & 0.70$\pm$0.2   & K5 & 14.51& 1.43$^{+0.16}_{-0.17}$   \\
    1313+4152     &  1525829295599805184 & 248$\pm$7    & 328$\pm2$        & 0.37$\pm$0.04 & 0.69$\pm$0.2   & K7 &16.18& 1.44$^{+0.19}_{-0.20}$   \\
    1527+3536     &  1375051479376039040 & 118$\pm$0.9  & 0.2556698$\pm2\cdot10^{-7}$       & 0.0\tablefootmark{5} & 0.62$\pm$0.01  & K9–M0 & 13.05& 0.98$\pm$0.03   \\
    1832–0119     &  4271998639836225920 & 181$\pm$2    & 545$\pm2$        & 0.44$\pm$0.05 & 0.63-1.00      & M2 & 15.62& 1.44$\pm$0.17   \\
    2128+3316     &  1854241667792418304 & 227$\pm$2    & 1430$\pm66$       & 0.59$\pm$0.02 & 0.70$\pm$0.2   & K5 & 14.87& 1.62$^{+0.17}_{-0.18}$   \\    
           \hline
        \end{tabular}
         \tablefoot{
                \tablefoottext{1}{The periods of 0616+2319 and 1527+3536 derive from the dedicated works of \citet{Yuan2023} and \citet{Lin2023} respectively. All other periods are from \citet{Andrews2022}.}
                \tablefoottext{2}{Mass of the main sequence star}
                \tablefoottext{3}{Spectral type of the main sequence star, from \citet{Yuan2023, Andrews2022, Lin2023, gaiadr3}}
                \tablefoottext{4}{Mass of the unseen companion}
                \tablefoottext{5}{\citet{Lin2023} used a circular Keplerian model for the orbital parameters fit}
                }
    \end{sidewaystable*}

\section{Observations and Data Analysis}
     
        \begin{table}[]
            \centering
            \footnotesize
            \caption{Journal of \textit{Swift} Observations}
            \label{table:2}
            \begin{tabular}{ccc}
            \hline\hline
            Source Name & Exp. Time & Obs. Date \\
                        & \multicolumn{1}{c}{s} & \\
            \hline
            0616+2319   & 5938 & 2023-10-31 - 2023-11-23\\
            1220+5841   & 4364 & 2023-03-08 - 2023-03-13\\ 
            1313+4152   & 4418 & 2023-03-13 - 2023-03-16\\ 
            1527+3536   & 3260 & 2022-10-30  \\
            1527+3536   & 5010 & 2022-11-20  \\
            1527+3536   & 9315 & 2023-01-01\\
            1832–0119   & 4309 & 2023-03-09\\
            2128+3316   & 4655 & 2023-03-15\\
            \hline 
            \end{tabular}
            
        \end{table}

Our targets were observed with \textit{Swift} through Target of Opportunity programs as summarised in Table\,\ref{table:2}, where we give the total exposure times obtained for the X-ray Telescope (XRT) (the observations  were  in some cases split over different epochs).  The XRT was in photon counting mode.  The data were processed using the XRTDAS package (version 5.7) distributed with HEASOFT (version 6.31), using the most recent release of the calibration database (CALDB version 10).  
    
Source counts for the spectral analysis of the only  source detected with the XRT (1527+3536) were extracted using a circular region of 30 pixels (70.8$\arcsec$), and the background was estimated using a region with the same shape and size and far from the source position. Spectral analysis was performed using the XSPEC package (version 12.13.0c). 
The upper limits for the non-detected sources were determined from the images in the 0.3-10 keV and 0.3-1 keV energy ranges using the Living Swift-XRT Point Source (LSXPS) catalog online upper limit server \citep{Evans2023,Evans2014}, which extracts counts from a 12 pixels region centered on the source position, estimates the background from the appropriate background map, and applies the \citet{Kraft1991} Bayesian method to estimate the upper limit.
 
All the targets were detected with the UV and Optical Telescope (UVOT), which operated with the filters UVW2 (central wavelength $\lambda$=192.8 nm and width $\delta\lambda$=65.7 nm) for two sources and with UVM2 ( $\lambda$= 224.6 nm, $\delta\lambda$=49.8 nm) for the other four sources (see Table\,\ref{table:3}). However, we note that the observations executed with the UVW2 filter are affected by contamination  from optical photons, caused by the red tail of the filter passband. For a K2V type star it is estimated that the UV portion is less than $\sim$20\% of the photon total flux \citep{Brown2010}, so any measured flux should be interpreted as an upper bound on the real flux.
    
 \section{Results}
        
In all the XRT observations pointed on 2MASS J152748.48$+$353657.2, a source was detected (S/N$>$10) at coordinates R.A. = 15$^\mathrm{h}$27$^\mathrm{m}$48$\fs$52, Dec. = +35$^{\circ}$36$^{\prime}$58$\farcs$8, with an uncertainty of 3.6" (90\% conf. level). This position is only 1.6" away from the \textit{Gaia} position of our target  and this is the only source in the error box of the ROSAT source 1RXS J152748.8+353658 (positional error 19\arcsec, 1-$\sigma$ radius). We are therefore confident that the source detected by the Swift/XRT is the candidate NS binary.

No evidence for variability was found by comparing the data of the three epochs (see Table\,\ref{table:4}). Therefore,  the subsequent analysis was carried out on the summed data, which provided a total of about 180 source counts in $\sim$18 ks of exposure. 
We extracted the source spectrum in the 0.3-10 keV energy range and fitted it applying the Cash statistics. 
Since all the models including an absorption component only provided an upper limit of $N_H < 10^{21}$ cm$^{-2}$ (90\% confidence level), and considering that the source is only 118 pc away, we performed the spectral analysis assuming there is no absorption. 
Good fits were obtained with either a  power-law model with a photon index of $\Gamma=1.9\pm0.2$, or a thermal plasma model (\textit{mekal}   in XSPEC) with temperature of $kT=3.5^{+1.5}_{-0.8}$\,keV.  In both cases, a flux of $\sim5-6\times10^{-13}$ erg~cm$^{-2}$~s$^{-1}$  in the 0.3-10 keV range was obtained. 
These values are consistent with the results reported by \citet{Mereghetti2022}, based only on the data of the first two epochs (see Table\,\ref{table:4}).
A fit with a black body model  was unacceptable (cstat/dof = 229.5/121). 
All the fit parameters are summarised in Table\,\ref{table:4}, where we also give the results obtained by fitting the individual spectra of the three epochs.
   
The other five sources were not detected by the XRT. The upper limits of the count rate in two energy ranges are given in  Table\,\ref{table:3}. The corresponding flux limits were computed assuming a power-law spectrum with index $\Gamma=2$ and no absorption. Luminosity upper limits were then computed using the source distances reported in Table \ref{table:3}.
    
We also derived upper limits on the temperature of a hypothetical black body emission, under the assumption of a spherical emitting surface with a radius of 10 km, at the distance of the systems. The values are reported in Table\,\ref{table:3}. 
       
 \section{Discussion}
      
\subsection{X-ray emission from 1527+3536  }
        
The nature of the X-ray emission from 1527+3536  was discussed by \citet{Lin2023} on the basis of its possible association with the source 1RXS J152748.8+353658 discovered in the ROSAT All Sky Survey. This association is now confirmed by our more precise localisation. 
The mass estimate of the compact object in this system obtained by \citet{Lin2023}, 0.98$\pm$0.03 \msun, is also compatible with a white dwarf (WD). We also note that \citet{Zhang2024}, using high resolution spectroscopy obtained with the Canada-Hawaii- France 3.6m Telescope, estimate a mass of 0.69$\pm$0.02, typical for a WD. However, \citet{Lin2023},  based on the X-ray luminosity being lower than that of intermediate polars and on the absence of dwarf nova-like outbursts in the long-term optical data, favor the NS interpretation. In this case, they claim that despite its binary nature, the source X-ray properties would resemble those of the so-called X-ray dim isolated NSs  (XDINS), a small class of nearby, isolated thermally emitting NSs  which might be the descendant of magnetars  \citep[see e.g.][]{Turolla2009}.

The Swift/XRT data presented here do not support the XDINS interpretation. In fact, all the isolated NSs of this class have very soft thermal spectra,  well fit by blackbody models with temperatures kT$<$100 eV. On the contrary, the XRT spectrum of 1527+3536 is rather hard and a BB model gives a bad fit, which might indicate a NS age greater than the range accepted for XDINs. Such a hard X-ray spectrum instead might be consistent with low level accretion from the companion star. Power-law components have been observed in neutron star low-mass X-ray binaries in quiescence, i.e., when accretion is mostly shut off, and interpreted as possibly due to some residual accretion ongoing \citep[e.g.][and references therein]{Degenaar2012}. In fact, \citet{Lin2023} interpret the presence of variable H$\alpha$ emission from this system as evidence for an accretion disk or stream.  The observed luminosity of $\sim10^{30}$  erg s$^{-1}$ requires an accretion rate of the order of $10^{10}$ g s$^{-1}$ if the compact object is a NS. In the case of a WD, the inferred accretion rate would lead to dwarf nova phenomena that are thus far not supported by long time-series data \citep{Lin2023}.
Although the late-type companion is not expected to provide such a rate through a strong stellar wind, we note that it might be filling (or close to fill) its Roche lobe which has a radius of only 0.66 \rsun.

Another possibility is that the X-ray emission is produced by a rotation-powered NS. In this case, assuming a typical efficiency of $\sim10^{-3}$ for conversion of rotational energy to X-rays, yields $\dot{E}_{rot} \sim10^{33-34}$  erg s$^{-1}$. In this case, the lack of a radio detection (with a deep upper limit of 5 $\mu$Jy at 1.2 GHz, \citet{Lin2023}) could be explained by an unfavorable beam direction. However, considering the small distance of only 100 pc, the absence of a bright Fermi/LAT $\gamma$-ray source at this position \citep{4LAT2020} makes this interpretation unlikely, considering that pulsar beams in $\gamma$-rays  are much wider than in the radio band \citep[see, e.g.,][]{Johnston2020}. 
Furthermore, no signs of irradiation from an energetic pulsar are visible in the optical data of the companion star. 

In summary, if the unseen component of this binary is a NS, we believe that low-level accretion from its companion is the most likely possibility to explain its X-ray emission. The alternative possibility of an accreting white dwarf, which would require a factor $\gtrsim500$ higher accretion rate, cannot be ruled out by the current X-ray data.
Finally, although the temperature derived with a mekal spectrum is rather high for a K9-M0 star, it cannot be excluded that the non collapsed component in this system be responsible for the X-ray emission (or at least part of it) if some magnetic coronal activity is present \citep{Preibisch2005, Drake2019}. Similar conclusions have also been reached for an analogous system, LAMOST J235456.73+335625.9, as shown by \citet{Zheng2023}.
\subsection{Upper limits}

Four of the systems were not detected by the XRT (1220+5841, 1832–0119, 1313+4152, 2128+3316), with very stringent upper limits on the X-ray luminosity, i.e., below 10$^{30}$ erg/s. They all have orbital periods longer than many months and, thus, very wide orbits.     
The optical components are low mass, late type stars that are clearly not filling their Roche lobes and do not have strong stellar winds. The candidate NSs in these binaries can thus be considered as virtually isolated objects for what concerns their X-ray emission.

Therefore, in addition to the processes discussed above for 1527+3536, we must also consider the possibility of X-ray emission powered by accretion from the interstellar medium (ISM).  
In a seminal paper, \citet{Ostriker1970} using the Bondi formulation  showed that if a NS accretes from the ISM, the typical expected luminosity L should be in soft X-rays, with 
    \begin{equation}
            L\sim 10^{32}~v_{10}^{-3}~n~ {\rm erg~ s^{-1}},\label{eq:1}
    \end{equation}
where $v_{10}$ is the velocity of the NS with respect to the medium in units of 10 km~s$^{-1}$, and $n$ is the ISM density in atoms~cm$^{-3}$.  \citet{Treves1991}  have shown that taking Eq. \ref{eq:1} at face value, a large number of isolated NSs were to appear in the ROSAT survey. 
Now it is clear that this is not the case, and the problem is rather to understand and interpret this absence. 
The explanation could be linked to a residual magnetic field, which somehow inhibits accretion, or to velocities that are higher than initially estimated \citep{Popov2023}.

The upper limits on the X-ray luminosity (Table \ref{table:3}) can be reconciled with  Eq. 1  assuming  a small ISM  
density, the impediment from the magnetic field,  very soft   emission peaking below the XRT band, or a moderate velocity of, e.g.,  $\sim$100 km/s (although such a velocity, typical of isolated NS, is unlikely for NSs bound in  binaries).

The case of 0616+2319 is slightly different, because its short orbital period of only $\sim21$ hours implies that this is a much more compact binary. However, the parameters derived by fitting its multi-wavelength light curves \citep{Yuan2023} indicate that also in this binary the non-collapsed component does not fill the Roche lobe.  Accretion from the companion's stellar wind might occur in this system, but the constraints on the mass accretion rate derived from our upper limit on the X-ray luminosity are not particularly useful, owing to the large distance of this source and the relative weakness of stellar winds from late G-type main sequence stars. 

\subsection{Search for UV excesses}
  
To compare our UV detections and upper limits with the available data for our targets at longer wavelengths, 
we searched the literature for photometric points using the catalogues available on the VizieR \citep{vizier} service. For each source, we queried for catalogue entries within 1\arcsec\ from the \textit{Gaia} position, and then inspected the resulting tables manually to check for any spurious association.  
We then plotted the resulting Spectral Energy Distributions (SED)   along with emission curves from a black body with temperature set on the basis of the spectral type of the companion star (as reported in Table \ref{table:1}) or (in the case of 1832-0119) based on the peak emission of the SED. 

In the case of 1527+3536, our UVOT flux  (Table 3 and Fig.1,2) agrees with the GALEX photometry reported by \citet{Lin2023} and therefore confirms the presence of a sizeable UV excess, which is apparent either comparing with a black body, or the stellar model used by \citet{Lin2023}.
In our data there is a possible indication of a 0.1 mag variability on month time scales (Tab.4). The excess could be due to the accretion process of the NS, as proposed by \citet{Lin2023}, but one cannot exclude some magnetic coronal activity of the non-collapsed component, as mentioned in the previous section, or emission from the surface of a WD with $T_{eff}\sim12000$ K.

For all the remaining sources, there is no evidence of UV emission in excess of the flux expected from the companion star. 

 \section{Conclusions}
        
We obtained Swift observations of a sample of  binary systems  selected from optical observations as  good candidates for hosting a NS or WD.
Only one system, 1527+3536, is detected in X-rays and in the UV. While this emission can be explained by accretion on a NS from its main sequence companion (almost) filling its Roche lobe, alternatives (such as a white dwarf, or X-ray emission from the main sequence star corona) cannot be completely ruled out.
   
In all other cases, no X-ray or UV excess emission is detected. Since the characteristics of these binaries make accretion from the non-compact object (through Roche lobe or stellar wind) unlikely, we tried to interpret the data within a XDINS-like scenario. In 4 cases (1220+5841, 1313+4152, 1832-0119, and 2128+3316) the upper limits we find require to invoke either a high velocity of the NS, small ISM density, accretion suppression by the magnetic field, or emission peaking in the very soft X-rays. For 0616+2319, the compactness of the system could allow for accretion from the main-sequence star stellar wind, but because of the larger distance of the system the constraints we can derive on the accretion rate are not very stringent.
   
Overall, all the systems we observed are compatible with the presence of a NS as the undetected companion, but we did not find any strong evidence of its presence, nor we can exclude their WD nature.
Observations in a softer X-ray band (performed with a detector whose sensitivity peaks below  the XRT energy band) or in the radio (to check for the presence of a pulsar-like activity) are the next natural steps in order to characterise these systems and fully assess the viability of using \textit{Gaia} data to select NS binary systems.
     
       \begin{table*}[]
            \centering
            \footnotesize
            \caption{Results\tablefootmark{1}}
            \label{table:3}
            \begin{tabular}{cccccclc}
            \hline\hline
            Source Name & 0.3–10 keV  rate & 0.3-1.0 keV  rate & 0.3–10 keV X-ray Flux\tablefootmark{2}  & X-ray Luminosity \tablefootmark{2} & BB Temp. & UVW2\tablefootmark{3,4} & UVM2\tablefootmark{3} \\
                        & \multicolumn{1}{c}{counts~s$^{-1}$}&\multicolumn{1}{c}{counts~s$^{-1}$}&\multicolumn{1}{c}{erg~cm$^{-2}$~s$^{-1}$} &\multicolumn{1}{c}{erg~s$^{-1}$} & \multicolumn{1}{c}{$10^5$ K} & \multicolumn{1}{c}{mag} & \multicolumn{1}{c}{mag} \\
            \hline
            0616+2319   &$<2.0\times 10^{-3}$&$<1.5\times 10^{-3}$& $<1.00\times 10^{-13}$        & $<1.48\times 10^{31}$ &$<$6.0 &  –  & 16.51$\pm$0.04              \\
            1220+5841   &$<2.5\times 10^{-3}$&$<2.5\times 10^{-3}$& $<1.25\times 10^{-13}$        & $<7.17\times 10^{29}$ &$<$4.2 & $>$20.0    & –             \\ 
            1313+4152   &$<1.6\times 10^{-3}$&$<1.6\times 10^{-3}$& $<8.00\times 10^{-14}$        & $<5.89\times 10^{29}$ &$<$4.1 & $>$21.5     & –              \\ 
            1527+3536   &$ 1.2\pm0.1\times 10^{-2}$&$6.2\pm0.7\times 10^{-3}$&  $ 5.20\pm0.7\times 10^{-13}$  & $8.0\pm1.0\times 10^{29}$     & – & – & 16.55$\pm$0.04              \\
            1832–0119   &$<2.5\times 10^{-3}$&$<1.6\times 10^{-3}$& $<1.25\times 10^{-13}$        & $<4.90\times 10^{29}$ &$<$4.0 & $>$19.04                & – \\
            2128+3316   &$<1.9\times 10^{-3}$&$<1.9\times 10^{-3}$& $<9.50\times 10^{-14}$        & $<5.86\times 10^{29}$ &$<$4.1 & $>$20.5                &  – \\
            \hline 
            \end{tabular}
            \tablefoot{
                \tablefoottext{1}{All errors and limits are quoted at 90\% confidence level.}
                \tablefoottext{2}{Flux and luminosity upper limits are calculated under the assumption of a power-law spectrum with index $\Gamma=2$.}
                \tablefoottext{3}{Vega system.}
                \tablefoottext{4}{Observations of late type stars with the UVW2 filter are affected by contamination from optical photons caused by the red tail of the filter passband. For a K2V type star it is estimated that the UV portion is less than only 20\% of the photons total flux \citet{Brown2010}, so any measured flux should be interpreted as an upper bound on the real flux.}
                }
        \end{table*}

        \begin{table*}[]
            \centering
            \caption{Spectral results for  1527+3536\tablefootmark{1}}
            \label{table:4}
            \footnotesize
            \begin{tabular}{ccllccllc}
            \hline\hline 
             Obs. Date & Exposure & \multicolumn{3}{c}{Power Law Model}   & \multicolumn{3}{c}{Thermal Plasma Model} & UVM2\tablefootmark{2} \\
                       &          &  \multicolumn{1}{c}{$\Gamma$}   & \multicolumn{1}{c}{Flux} & \multicolumn{1}{c}{cstat/dof}     & \multicolumn{1}{c}{kT} & \multicolumn{1}{c}{Flux} & \multicolumn{1}{c}{cstat/dof} &  \\
                       &  \multicolumn{1}{c}{$s$} &    & \multicolumn{1}{c}{erg~cm$^{-2}$~s$^{-1}$}&  & keV  & \multicolumn{1}{c}{erg~cm$^{-2}$~s$^{-1}$} & & \multicolumn{1}{c}{mag} \\
            \hline
            2022-10-30      & 3260  & $2.0\pm0.5$ & $3.4^{+1.6}_{-1.1}\times 10^{-13}$    & 30.84/28     & $2.80^{+4.4}_{-0.8}$       & $(2.4\pm1.7)\times 10^{-13}$ & 30.23/28    & $16.51\pm0.04$  \\
            2022-11-20      & 5010  & $2.3\pm0.4$ & $(3.6\pm0.9)\times 10^{-13}$            & 45.07/47     & $2.3\pm1.0 $              & $(1.4\pm0.9)\times 10^{-13}$ & 46.63/47    & $16.43\pm0.04$   \\
            2023-01-01      & 9315  & $1.8\pm0.3$ & $(4.0\pm0.9)\times 10^{-13}$            & 72.8/76     & $3.5^{+2.8}_{-1.2}$       & $(3.5\pm0.8)\times 10^{-13}$ & 74.1/76    & $16.55\pm0.04$   \\
            \hline
            \multicolumn{9}{c}{Global Fit}\\
                            & 17585 & $1.9\pm0.2$ & $(5.2\pm0.7)\times 10^{-13}$           & 124.92/121   & $3.5^{+1.5}_{-0.8}$       & $(4.9\pm0.7)\times 10^{-13}$ & 136.01/121 & $16.51\pm0.04$  \\
            \hline
            \end{tabular} 
         \tablefoot{
           \tablefoottext{1}{All errors are quoted at 90\% confidence level.}
           \tablefoottext{2}{Vega system.}
          }   
        \end{table*}

 \begin{acknowledgements}
This research has made use of the VizieR catalogue access tool, CDS, Strasbourg, France (DOI : 10.26093/cds/vizier). The original description of the VizieR service was published in 2000, A\&AS 143, 23.
This work made use of data supplied by the UK Swift Science Data Centre at the University of Leicester. AM, FCZ and NR are supported by the H2020 ERC Consolidator Grant “MAGNESIA” under grant agreement No. 817661 (PI: Rea), grant SGR2021-01269 (PI: Rea), and partially supported by the program Unidad de Excelencia Mar\'ia de Maeztu CEX2020-001058-M. FCZ is also supported by a Ram\'on y Cajal fellowship (grant agreement RYC2021-030888-I).
This work has made use of data from the European Space Agency (ESA) mission
{\it Gaia} (\url{https://www.cosmos.esa.int/gaia}), processed by the {\it Gaia}
Data Processing and Analysis Consortium (DPAC,
\url{https://www.cosmos.esa.int/web/gaia/dpac/consortium}). Funding for the DPAC
has been provided by national institutions, in particular the institutions
participating in the {\it Gaia} Multilateral Agreement.
\end{acknowledgements}

   \bibliographystyle{aa} 
   \bibliography{xdim_ns.bib} 

\newpage
\newpage

    \begin{figure*}
        \centering
            \includegraphics[width=1.0\columnwidth]{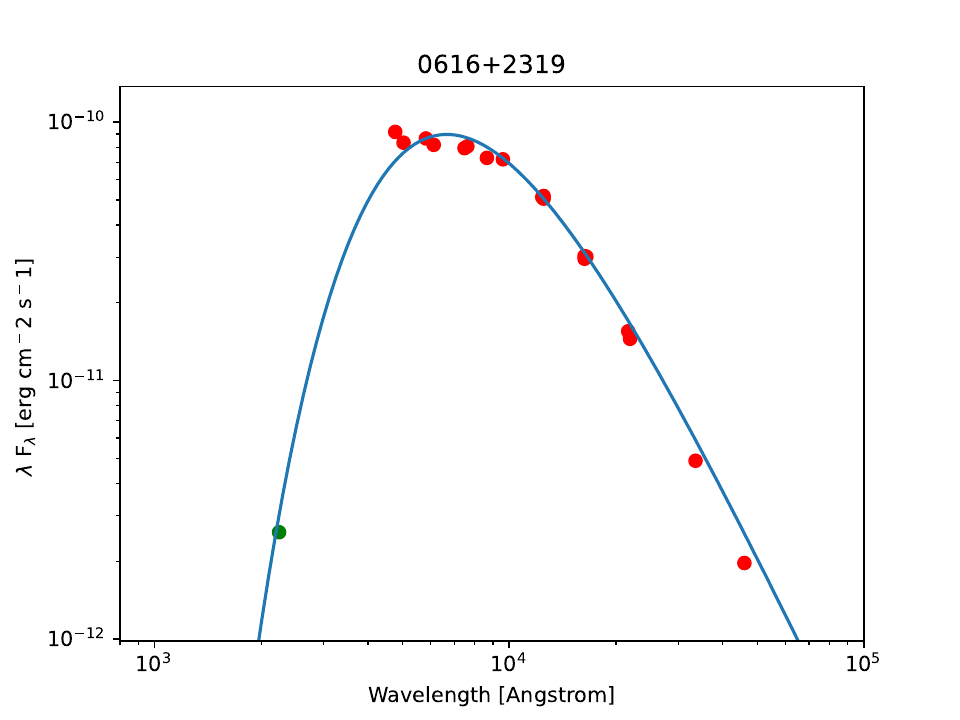}
            \includegraphics[width=1.0\columnwidth]{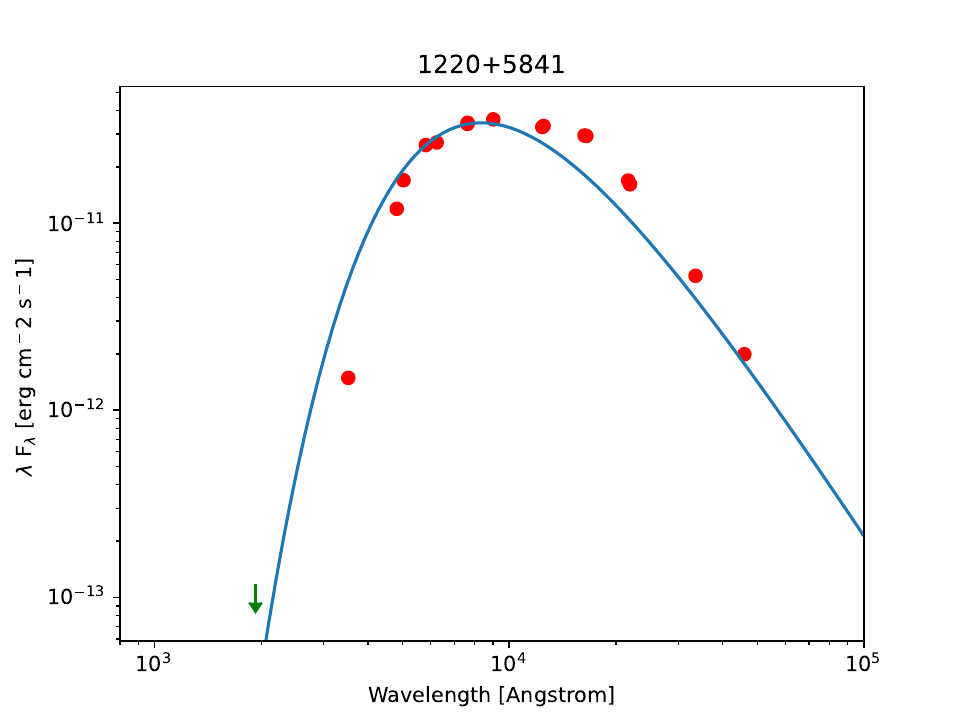}
            \caption{SED for the systems presented on this study. Red points are literature data from Vizier (GALEX, {\it Gaia}, SDSS, 2MASS, PAN-STARRS, WISE data). Green points are Swift UVOT data from our ToO campaign, Blue lines are blackbody emission profiles with temperatures determined by the main sequence star companion stellar type, normalized to the {\it Gaia} G flux.}
            \label{Fig:1}
    \end{figure*}
    \addtocounter{figure}{-1}

    \begin{figure*}
        \centering
            \includegraphics[width=1.0\columnwidth]{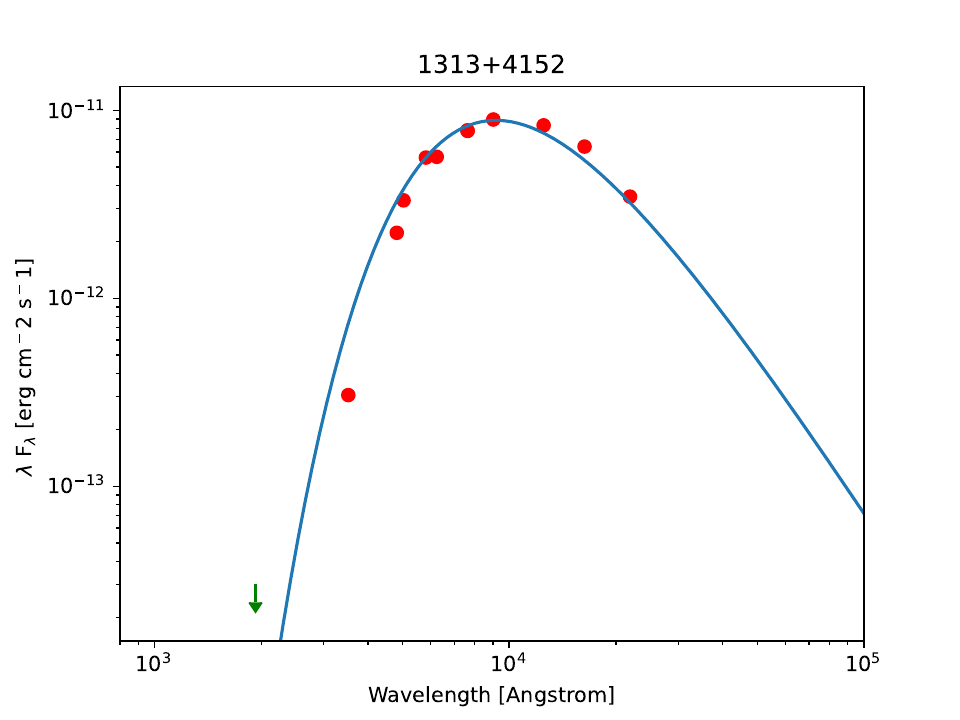}
            \includegraphics[width=1.0\columnwidth]{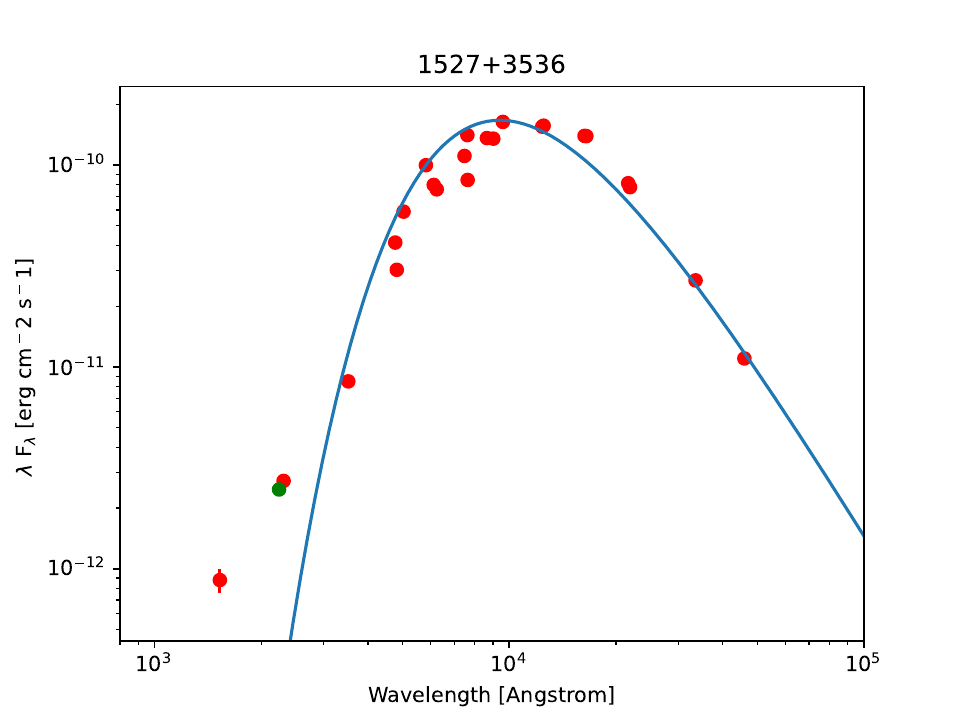}
            \caption{continued}
            \label{Fig:1}
    \end{figure*}
    \addtocounter{figure}{-1}

    \begin{figure*}
        \centering
            \includegraphics[width=1.0\columnwidth]{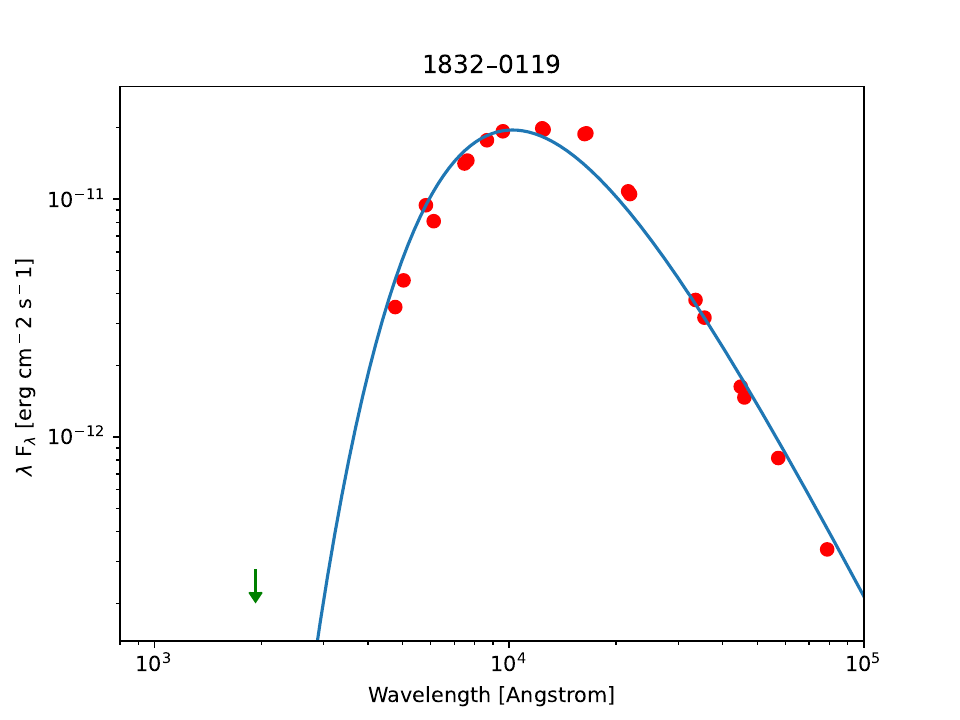}
            \includegraphics[width=1.0\columnwidth]{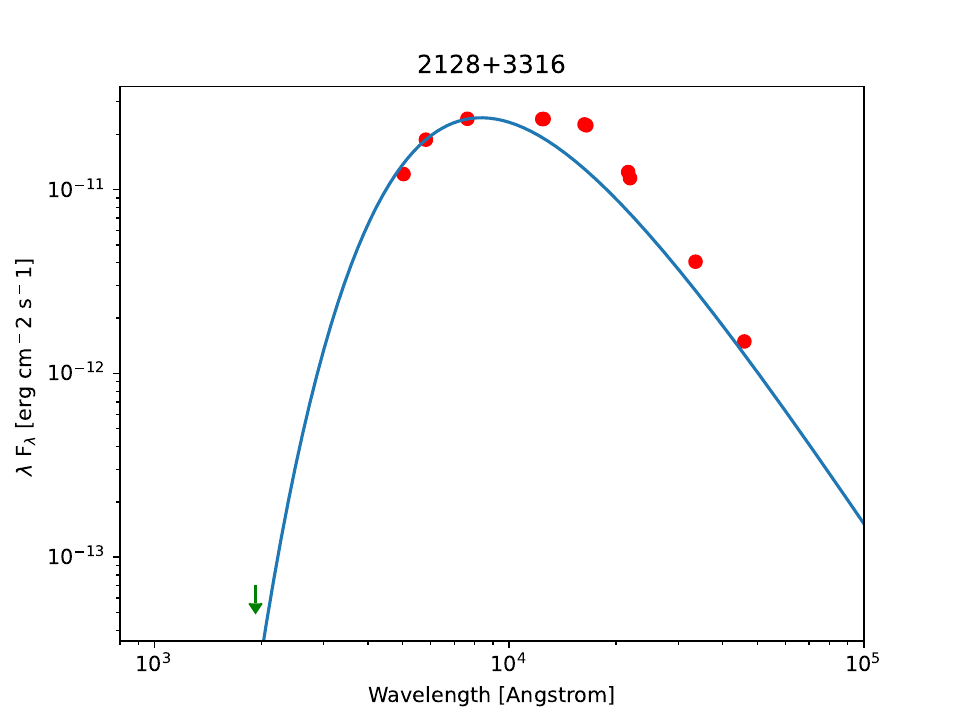}
            \caption{continued}
            \label{Fig:1}
    \end{figure*}
    
    \begin{figure*}
        \centering
            \includegraphics[width=1.0\columnwidth]{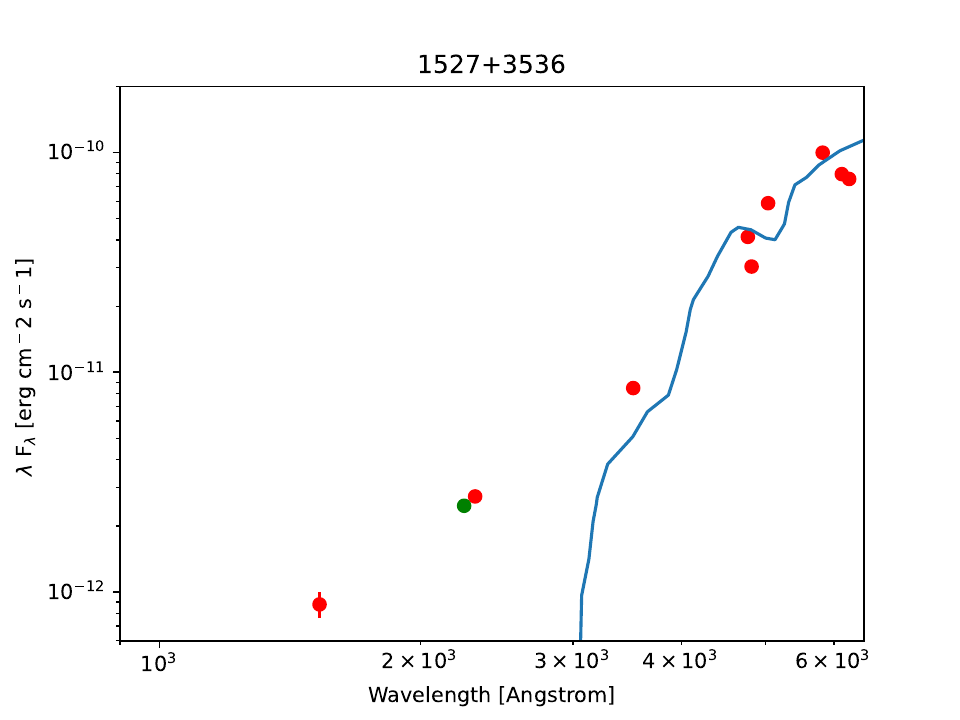}\\
            \caption{SED for 1527+3536. Red points are literature data from Vizier (GALEX, {\it Gaia}, SDSS, 2MASS, PAN-STARRS, WISE data). The green point is Swift UVOT data from our ToO campaign, the blue lines is the main sequence companion star spectral fit derived from \citet{Lin2023}.}
        \label{Fig:2}
    \end{figure*}

\end{document}